\documentclass[conference]{IEEEtran}
\IEEEoverridecommandlockouts

\usepackage{graphicx}
\usepackage{algorithm}
\usepackage{algorithmic}
\usepackage{subcaption}
\usepackage{cite}
\usepackage{url}
\usepackage{mdwmath}     
\usepackage{mdwtab}
\usepackage{amsfonts,amsmath,color,amssymb,amsxtra,amsbsy,floatflt}
\usepackage{indentfirst} 
\usepackage{tabularx}
\usepackage{booktabs}
\usepackage{multirow}
\usepackage{nth}
\usepackage{comment}
\usepackage{epstopdf}
\usepackage{bbold}
\usepackage{multirow}
\usepackage{hhline}

\newcommand{\name}{NSBchain}

\begin{document}
\graphicspath{{./figs/}}

\title{\name{}: A Secure Blockchain Framework for Network Slicing Brokerage}

\author{
    \IEEEauthorblockN{Lanfranco Zanzi\IEEEauthorrefmark{1}\IEEEauthorrefmark{2}, Antonio Albanese\IEEEauthorrefmark{1}\IEEEauthorrefmark{3},
    Vincenzo Sciancalepore\IEEEauthorrefmark{1}, Xavier Costa-P\'erez\IEEEauthorrefmark{1}}
    \IEEEauthorblockA{
	\IEEEauthorrefmark{1}NEC Laboratories Europe, Heidelberg, Germany\\
    \IEEEauthorrefmark{2}Technische Universit\"at Kaiserslautern, Kaiserslautern, Germany\\
 	\IEEEauthorrefmark{3}Universidad Carlos III de Madrid, Madrid, Spain
 	\thanks{\textsuperscript{*}Corresponding e-mails: \{name.surname\}@neclab.eu}}
}

\maketitle

\setlength{\textfloatsep}{5pt}

\begin{abstract}
With the advent of revolutionary technologies, such as virtualization and softwarization, a novel concept for 5G networks and beyond has been unveiled: \emph{Network Slicing}. Initially driven by the research community, standardization bodies as 3GPP have embraced it as a promising solution to revolutionize the traditional mobile telecommunication market by enabling new business models opportunities. \emph{Network Slicing} is envisioned to open up the telecom market to new players such as \emph{Industry Verticals}, e.g. automotive, smart factories, e-health, etc. Given the large number of potential new business players, dubbed as network tenants, novel solutions are required to  accommodate their needs in a cost-efficient and secure manner.


In this paper, we propose \name{}, a novel \emph{network slicing brokering (NSB)} solution, which leverages on the widely adopted \emph{Blockchain} technology to address the new business models needs beyond traditional network sharing agreements. \name{} defines a new entity, the \emph{Intermediate Broker (IB)}, which enables \emph{Infrastructure Providers (InPs)} to allocate network resources to IBs through smart contracts and IBs to assign and re-distribute their resources among tenants in a \emph{secure, automated and scalable} manner. 
We conducted an extensive performance evaluation by means of an open-source blockchain platform that proves the feasibility of our proposed framework considering a large number of tenants and two different consensus algorithms.

\end{abstract}


\section{Introduction}
\label{sec:introduction}

With the announced arrival of the $5^{th}$ generation of mobile networks (5G), vertical industries can embrace the mobile ecosystem and explore novel sources of revenues. Overcoming the traditional telecom stagnation on connectivity services, \emph{network slicing} expands telecom services towards dedicated virtual network instances, or slices, customized to meet specific industry verticals service requirements.
The advantage coming from the adoption of the network slicing paradigm is two-fold: $i$) Infrastructure Providers (InPs) will be able to reach greater levels of resource sharing, thus increasing the actual utilization of their physical infrastructure by means of statistical multiplexing of requests coming from $3^\textit{-rd}$ party vertical industries~\cite{StatisticalMultiplexing};~$ii$) vertical industries will benefit from dedicated mobile network slices enabling advanced  services to their final users, with specific Quality of Service (QoS) and Service Level Agreements (SLAs).

The novel network slicing paradigm, made available by the latest developments on virtualization and softwarization technologies, enables advanced and dynamic resource allocation schemes built on top of modular mobile architectures and commoditized platforms. Such advanced resource allocation mechanisms must deal with a heterogeneous and wide set of vertical requirements to satisfy per-slice  performance guarantees. In this context, the figure of the \emph{Network Slicing Broker (NSB)}, firstly introduced in~\cite{samdanis2016network}, acts as an entity in charge of mediating between industry verticals' slice requests and the mobile infrastructure resource orchestrator. 

In this paper, we extend the NSB concept towards further dividing the value chain and allowing the entrance of new players in a similar manner as mobile virtual network operators (MVNOs) did in telecom networks. MVNOs allowed InPs to address specific market niches, which they did not manage to tap into due to the \emph{subscriber acquisition costs}. The new challenge here is that, while the number of MVNOs is rather small in established mobile markets, network slicing is expected to accommodate hundreds to thousands of new industry vertical tenants, ranging from full coverage connected car platforms to localized IoT deployments.

In order to achieve this, we introduce the figure of the \emph{Intermediate Broker (IB)}, which leverages on \emph{Blockchain} technology to develop the \emph{network slicing brokering (NSB)} solution \name{}, enabling \emph{Infrastructure Providers (InPs)} to allocate network resources to IBs through smart contracts and IBs to allocate and re-distribute their resources among tenants in a \emph{secure, automated and scalable} manner. While MVNO agreements with InPs have to go through a regular \emph{offline contract signature} process, \name{} enables a much faster, scalable and cost-efficient \emph{secure online digital signature} process for the resource allocation transactions.

The contributions of this paper can be summarized as follows:
\begin{itemize}
    \item Introduction of a novel hierarchical network slicing brokering framework based on blockchain to support the evolution of the telecom business model.
    \item Design of a Blockchain-based smart contracts and consensus system that allows, in sliced networks, dynamic resource exchange among tenants.
    \item Evaluation of our \name{} framework building on top of the HyperLedger platform \cite{hyperledger} and analysis of key performance features, e.g. transaction throughput, communication latency and platform scalability.
\end{itemize}

The remainder of this paper is as follows.
Section~\ref{sec:background} provides an overview of the network slicing concept and summarizes  the related work in the field.
Section~\ref{sec:hier_architecture} introduces the conceptual architecture and means to support network slicing in 5G systems, proposing the figure of the \emph{Intermediate Broker} and  discussing the applicability of the blockchain technology in the network slice resource brokering scenario.
Section~\ref{sec:proposed} proposes the novel \name{} framework to 
support the interaction between stakeholders as well as automatize the transfer of resource ownership in a secure way,
whereas Section~\ref{sec:perf_eval} validates the overall platform presenting a Proof-of-Concept implementation and its performance. Finally, Section~\ref{sec:conclusion} concludes the paper.

\section{The Network Slicing Brokerage Process}
\label{sec:background}

The network slicing business model revolves around three main entities~\cite{samdanis2016network}:
\textit{i)} the~\emph{Infrastructure Provider (InP)}, which is the owner of the mobile network physical infrastructure and responsible for its maintenance,
\textit{ii)} the~\emph{Network Slice Tenants}, which are those business entities, e.g., Over-The-Top (OTT) service providers or $3^{\textit{rd}}$-party vertical industries, interested in renting a slice of the mobile network from the InP to provide tailored services to their customers through allocation of dedicated resources,
\textit{iii)} the~\emph{Network Slice Broker (NSB)}, which is in charge of mediating between tenants' requirements and network resource availability, and instructing the physical infrastructure to accommodate requests.

\begin{figure}[t!]
      \centering
      \includegraphics[clip, trim= 1cm 5cm 0cm 0cm,   width=\linewidth]{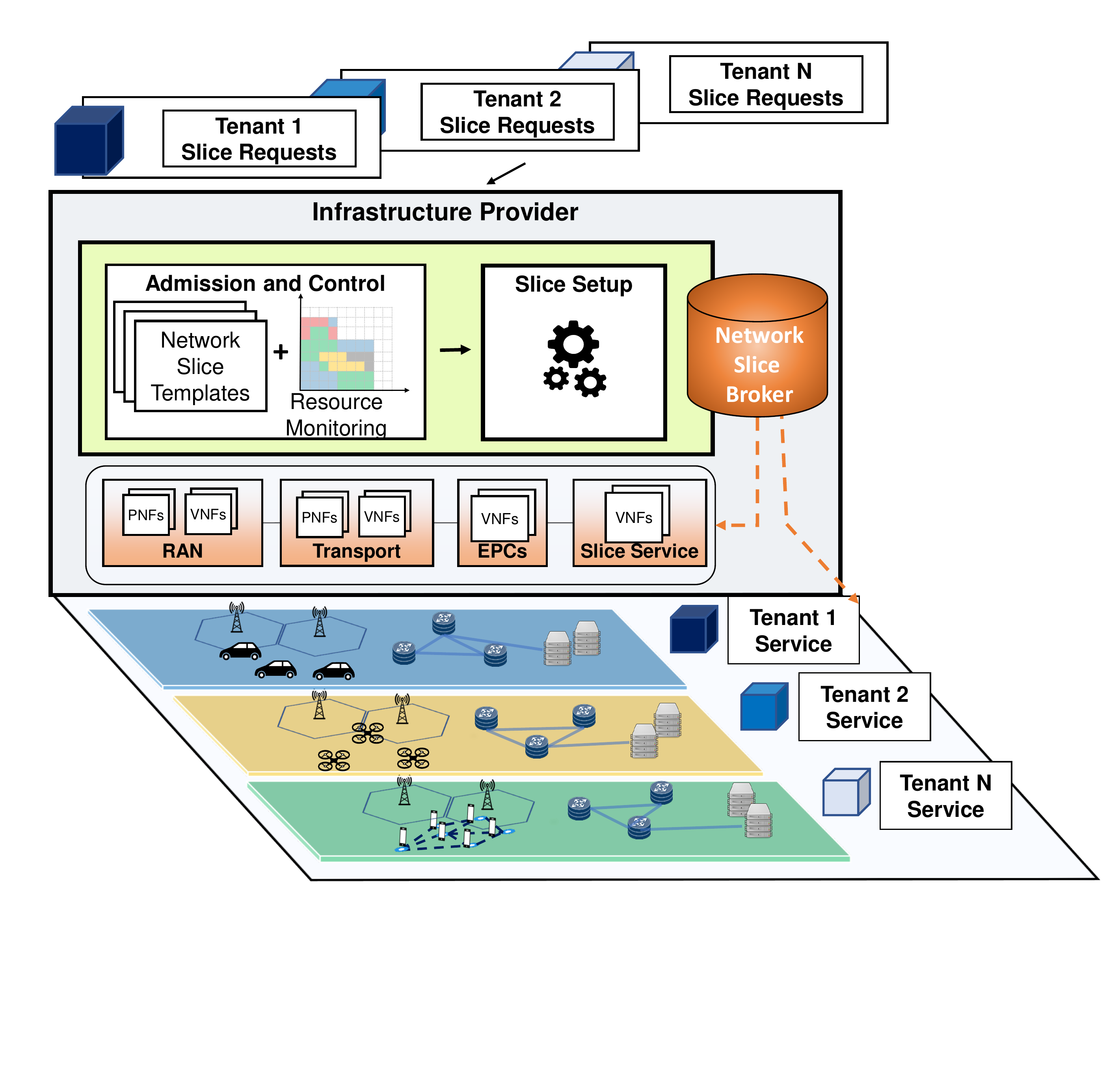}
      \caption{\small Network slice brokerage overview}
        \label{fig:architecture}
\end{figure}

In more detail, upon slice requests arrivals, the NSB is in charge of running an admission  control mechanism, and if granted, deploying the new slices in the system. Such admission control mechanism involves the evaluation of the slice resource requirements against the resource availability over the different network domains, Radio Access Network (RAN), transport, and core. Keeping running slices SLAs isolated from newcomers is of paramount importance in this scenario as it shall avoid resource shortage that might impact the service delivery. As different tenants may require a diverse set of network resources, the admissibility of each slice request depends on an elaborated multi-domain optimization problem, see for instance~\cite{Slicing2018CoNEXT}. To ease this task, a common solution accounts for the usage of a predefined set of Network Slice Templates (NSTs)\cite{TS28.531}. Each template specifies static parameters and functional components of different network slice types as well as the relevant attribute's value in terms of resource allocation requirements necessary to satisfy the service provisioning.
An illustration of the workflow is depicted in Fig.~\ref{fig:architecture}.

\subsection{Related Work}
\label{subsec:related}

Both network slicing and blockchain paradigms recently attracted wide interest as a consequence of the hype around 5G mobile networks and cryptocurrencies. Therefore, perhaps not surprisingly, several research efforts started investigating how to combine these two emerging technologies.

In~\cite{NSB_5G} the authors present a study on the leasing ledger concept proposing the blockchain technology as a means to overcome absence of trust in data management and satisfy the need for automated solutions in industrial network facilities.
One of the key-features inherited by the blockchain technology is indeed the capability of providing trust in a distributed way.
The authors of~\cite{Rawat2018} exploit this feature to minimize (discourage) over-committing issues during the negotiation of SLAs and radio frequency channels assignments between InP and Mobile Virtual Network Operators (MVNOs). Differently from our work, they only focus on RAN-specific resource negotiation without considering other network domains. Recent related work along our proposal has been presented by~\cite{VehicularBlockchain} in the context of vehicular ad-hoc network communication. Despite a complete analysis about security and performance aspects, no guidance is provided regarding functions and/or consensus protocols to achieve the complete solution.
Finally,~\cite{Rebello2019} proposes to extend the NFV-MANO architecture to account for a dedicated API through which network slices can be configured and orchestrated according to the negotiated transactions. As future work,~\cite{Rebello2019} highlights the need for a consensus algorithm able to manage, in an efficient manner, the huge number of interactions expected in slicing systems.

The key novelty of our framework is the capability to support the network slice resource brokerage process in an end-to-end fashion, embracing the multi-domain nature of the network slicing paradigm and its need to guarantee heterogeneous tenants' requirements, even at fine-grained granularity. Conversely, none of these prior works fully investigates a multi-tenant multi-domain scenario, limiting their analysis at domain-specific implementations.



\section{The Hierarchical Architecture to evolve the Network Slicing Market}
\label{sec:hier_architecture}

While the network slice market is envisioned to unlock a wide set of business opportunities\cite{AdLittle}, the management of a multitude of relatively small network slices---if geographically constrained like small business industries and factories---introduce additional complexity in the orchestration process when performed in a centralized fashion: operators might not want to undertake it.

The network slice ecosystem is envisioned to support dynamic and real-time resource allocation over the mobile network. In such a fast-changing scenario, tenant requirements may vary as a result of external causes, e.g., end-users' mobility, possibly leaving tenants with under or over-provisioned network slices and the need of acquiring/releasing resources.

\begin{figure}[t!]
      \centering
      \includegraphics[clip, width=\linewidth, trim = 2cm 4.5cm 1cm 2.5cm]{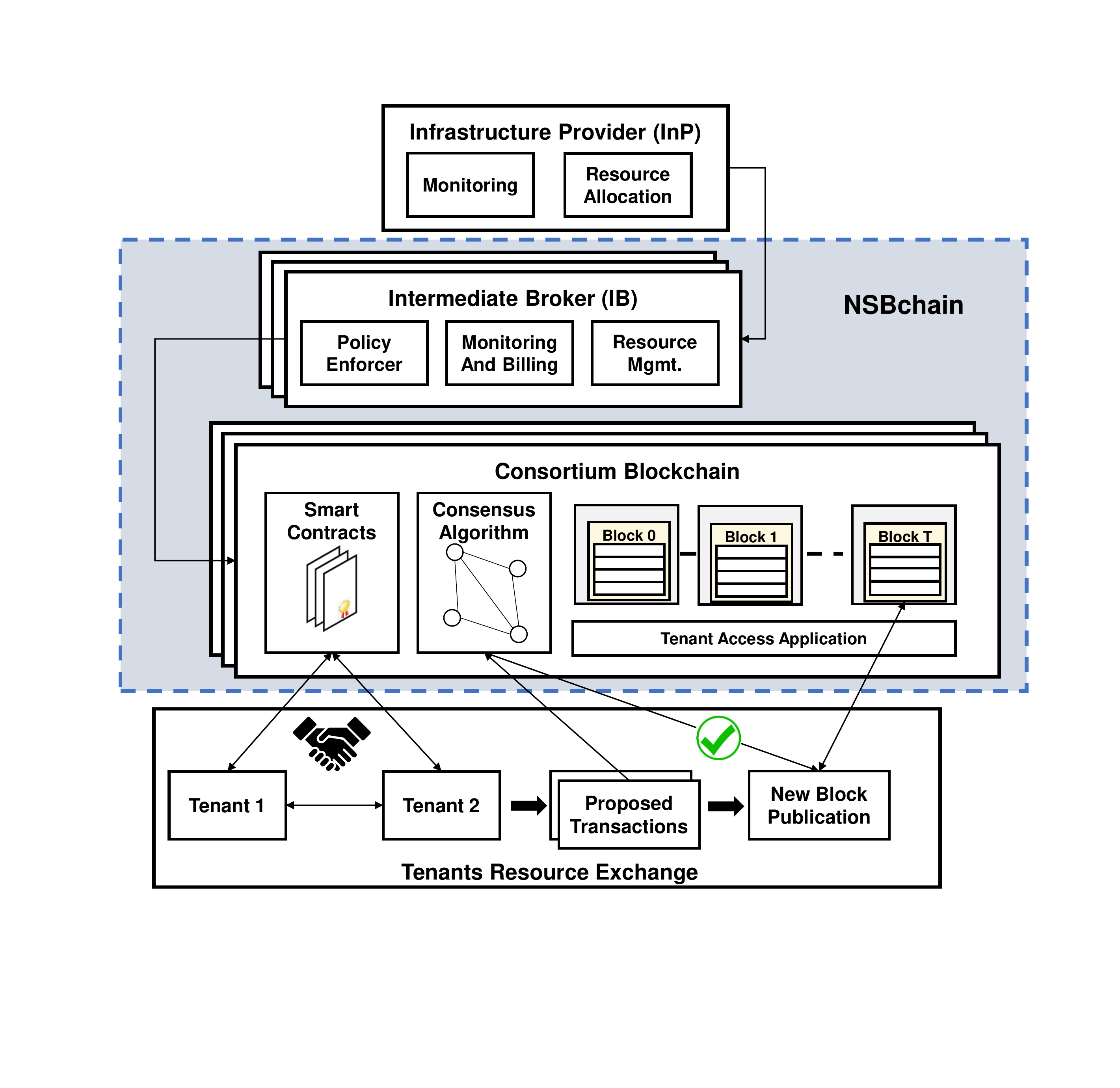}
      \caption{\small A distributed hierarchical architecture for network slicing.}
        \label{fig:distributed_architecture}       
\end{figure}

In this context, the roles of the InP and the wholesaler can be comparable. From this perspective, it is preferable to deal with the exchange of big quantities of goods to intermediate retailers rather than trading, with a significant increase of management costs, small quantities directly with the end-users. Thus, this opens up to new marketing opportunities for $3^{\textit{rd}}$-party entities willing to play the role of retailers, e.g., Mobile Virtual Network Operators, municipalities in case of public events, highway operators and factories, which may buy a quota of network resources from the InP and re-sell it to final tenants. We define such business entities as \emph{Intermediate Brokers} (IBs). 
We envision the network slicing economy as an open market where tenants can select the IB that best suits their requirements, e.g., better price, thus leading to the creation of consortia of tenants under the management of the same IB.
The proposed architecture is depicted in Fig.~\ref{fig:distributed_architecture}.

{\bf Technical challenges.}
In order to support the hierarchical structure above-described as well as the additional management and security complexity inherited by this enhanced business model, several challenges must be considered: flexibility and scalability are key-features for next generation mobile networks. 

Network slices should meet tailored and fast service provisioning requirements as dictated by the service diversity foreseen in the 5G era. Fully automated solutions are thus necessary to keep efficient network operations and management while reducing costs. End-user mobility aspects and interference management bring additional complexity in the network slicing context, especially for real-time use-cases. The assignment of network resources to tenants in such cases requires the resource allocation process to evolve dynamically following tenant demand variations.

At the same time, the chain of network resource loans must be negotiated in a secure, transparent and fast way~\cite{PermissionlessSurvey,BSecNFVO}, such that the lifecycle of each slice is not affected. Current mobile network sharing solutions require long negotiation processes that hardly fit within short time-to-market deployments of the 5G use-cases.

Due to its decentralized nature, the blockchain technology well suits these requirements. The distributed ledger allows all members of the system to be aware of the current (and past) network resource availability as well as to be informed, in real-time, about the dynamic exchange of resources through a public hash-chain of blocks provided with valid transactions. A secure resource exchange is guaranteed by smart contracts and distributed consensus algorithms, allowing the system to evolve autonomously without the need of centralized authorities.

\subsection{Blockchain}
\label{subsec:blockchain}
\begin{algorithm}[t]
\caption{Smart Contract implementing an auction-based resource allocation scheme.}
\label{algorithm}
\algsetup{indent=2em}
\begin{footnotesize}

    \begin{algorithmic}[1]
     \STATE {\bfseries Input:} AuctionEndTime, ResourceSet
     \STATE {\bfseries Initialize:} HighestBidderID = 0x00, HighestBid = 0
       \WHILE{Now() $\leq$ AuctionEndTime}
            \STATE ListOfBids = CollectBids();
            \FOR{CurrentBid $\in$ ListOfBids}
            \IF{CurrentBid.value $>$ HighestBid}
            \STATE HighestBidderID = CurrentBid.peerID;
            \ENDIF
            \ENDFOR
       \ENDWHILE
      \STATE Notify(HighestBidderID, "Your Bid was the highest.");
      \STATE Assign(ResourceSet, HighestBidderID);

\end{algorithmic}
\end{footnotesize}
\end{algorithm}

Despite becoming famous for the hype around cryptocurrencies, the blockchain technology applicability is not limited to that scope. In its simplest definition, a blockchain is a distributed data structure shared among the members of the network. Each block stores information about a set of transactions e.g., timestamp, amount of good exchanged, partners involved and most importantly a reference to the previous block of the chain (usually the hash of its content). The creation of new blocks involves secure cryptographic mechanisms that make the chain unalterable and safe against fraudulent attacks. The content of a blockchain database is broadcast and updated in a decentralized manner, being the absence of a centralized control an advantage for data transparency. 

However, the decentralized architecture implies synchronization issues, for example, when dealing with the insertion of new blocks in the chain. This calls for the introduction of a \emph{consensus} mechanism to keep the information contained in the ledger coherent within the network. Several algorithms are available in the literature showing advantages and drawbacks, e.g., proof of elapsed time, proof of work, and so on~\cite{SurveyConsensus}.

We can identify two types of blockchain: $i$) \emph{Permissionless} chains allow anyone to read, to write and to participate in the creation of the ledger, $ii$) \emph{Permissioned} blockchains pose restriction on who is allowed to participate in the network activities, e.g., limiting the kind of transactions.
Considering the enterprise facility represented by a mobile network infrastructure, permissioned access is preferable to maintain high security levels. To this aim, permissioned blockchains often exploit Trusted Execution Environments (TEE) to securely onboard participants and assist with the establishment of the consortium that composes the blockchain network. Such scheme also avoids the need of energy consuming activities related to block validation process, which has been identified as one of the main drawbacks of public blockchain systems~\cite{Energy}. Therefore, we can assume that peer nodes admitted in the system are not malicious and rational, i.e., profit driven.

In the blockchain context, smart contracts are often used to automatize the exchange of goods in reply to trigger events. A smart contract can be defined as an agent that translates contractual clauses into self-enforcing software that minimizes the need of trusted intermediaries. SCs are stored in the blockchain and provided with a unique address, making it easy to be reached from all the peers in the network and inheriting useful security features like distributed consensus agreements to prevent fraudulent usages.
The implementation of smart contracts often implies the usage of high-level programming languages, which are then compiled into low-level byte-coded languages and loaded into the blockchain to ensure immutability.

In our framework, we exploit SCs to guarantee reliable auditing and enforce IB-specific policies in the management of requests. For example, one IB may decide to auction his share of resources in different ways~\cite{OnlineAuction,Auction_ICC} or simply sell them to the first coming tenant. The pseudocode of an auction-based SC implementation is shown in Algorithm~\ref{algorithm}.
Peer nodes can invoke a SC by sending transactions to its address. In more detail, if a new transaction is proposed in the system, the contract address can be inserted as recipient address of the transaction. To validate the resource exchange, all the peer entities execute the code using, among the others, transaction payloads and current system state as input arguments of the call~\cite{Luu_SmartContract}. The participation in the consensus protocol finally assures that the new output ledger comes from valid transactions.
\begin{figure}[t!]
      \centering
      \includegraphics[clip, trim= 5cm 11cm 5cm 8cm,   width=0.9\linewidth]{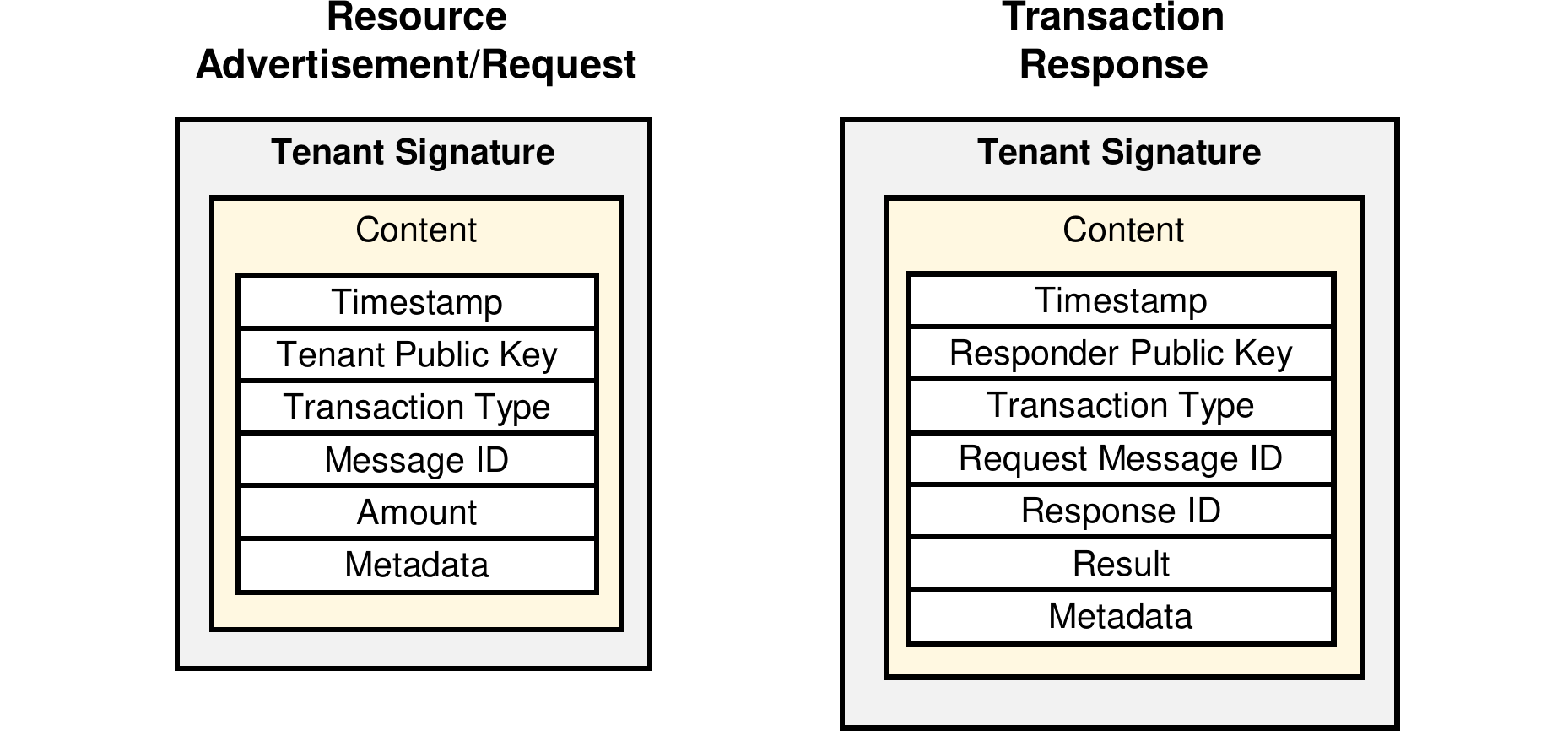}
      \caption{\small Example of transaction message exchange within \name{}. }
        \label{fig:message_structure}
\end{figure}

\section{The \name{} framework}
\label{sec:proposed}


Hereafter, we introduce our novel framework, namely \name{}, showing the main advantages and limitations when implemented in real deployments.

\subsection{Smart Contracts}
Analytically, let us introduce $\mathcal{B} = \{ b_1, b_k, \dots ,b_K \}$ as the set of IBs allowed to trade network resources, and $\mathcal{T}_k = \{ \tau_1, \tau_t, \dots ,\tau_T  \}$ as the set of tenants admitted within the consortium of IB $b_k$.
Being a permission-based system, our framework requires an invitation for participation\footnote{While the admission procedure is out of the scope of this paper, it is assumed that such a mechanism is in place and managed by the InP to guarantee that only trustworthy entities are admitted.}. To guarantee secure message exchange, each entity is provided with a cryptographic key pair $\{K_{priv},K_{pub}\}$. The usage of group signature schemes and the generation of new key pair for every message exchange is preferable to avoid reply attacks~\cite{HomeChain}.

We detail in the following the main steps involved in the creation and management network slices on a blockchain-based platform providing a mathematical background for the consensus process and the overall revenue maximization.
\indent \textbf{System Setup}. In order to enable dynamic resource exchange among tenants, a dedicated blockchain must be set up for each consortium of tenants. Each IB $b_k$ deploys the first block of the chain and loads a registry of resources $\mathcal{R}_k = \{ r_1, r_i, \dots ,r_I \}$ into such a block, which reflects the amount of $i$-type resources, with $i\in\mathcal{I}$, originally assigned by the InP. This step is required to avoid over-selling so as to limit the availability of resources in the blockchain.
Each IB $b_k$ can define leasing policies and code them into a set of SCs, which are then available to all tenants in the consortium. Finally, each IB $b_k$ is in charge of assigning the initial share of resources to admitted tenants. 

\indent \textbf{Message Exchange}. Upon private exchange domain creation, network slice requests can be dispatched among the network of peers.
According to their real-time requirements, tenants may decide to publish a resource advertisement or a resource request message. In the former case, the current owner of resources decides to release some of his shares making them available on the market. In the latter case, the tenant broadcasts its need to other tenants, which may be interested in providing their quota. To guarantee authentication, each message is signed with the sender private key and uniquely identified by an ID number. A simplified message structure is depicted in Fig.~\ref{fig:message_structure}.

The network slicing brokerage must deal with multi-domain resource allocation problems. In its simplest definition, a resource request from tenant $\tau$ can be defined as a tuple $\Psi_\tau = [\pi_1^{(\tau)}, \pi_i^{(\tau)}, \dots , \pi_I^{(\tau)} | \theta_1^{(\tau)}, \dots, \theta_I^{(\tau)} ]$, where $\pi_i^{(\tau)}$ represents the required amount of $i$-type resources, and $\theta_i^{(\tau)}$ is the price to be paid. It should be noticed that we do not pose any limitation on the nature of exchanged resources, and that the proposed resource request scheme easily accommodates heterogeneous resource specifications. For example, a tenant could be more interested in trading only Radio Access Network (RAN) resources at the edge of the network, e.g., for delay sensitive applications, while others may be more interested in cloud resources, e.g., storage and processing power for data analytic applications in the context of the Internet of Things (IoT).

\indent \textbf{Billing Management}. Interestingly, a blockchain can be viewed as a transaction-based state machine, wherein its state is updated every time consensus is reached on a set of transactions. To this aim, orderer nodes can be introduced and exploited to collect and sort proposed transactions by arrival time. Such nodes are usually not involved in the validation process, however they may allow decoupling and parallel processing of ordering and validation functionalities thereby improving the overall system efficiency~\cite{Rebello2019}. 

\begin{figure}[t!]
      \centering
      \includegraphics[clip, trim= 2.5cm 5cm 2.5cm 2cm,   width=0.9\linewidth]{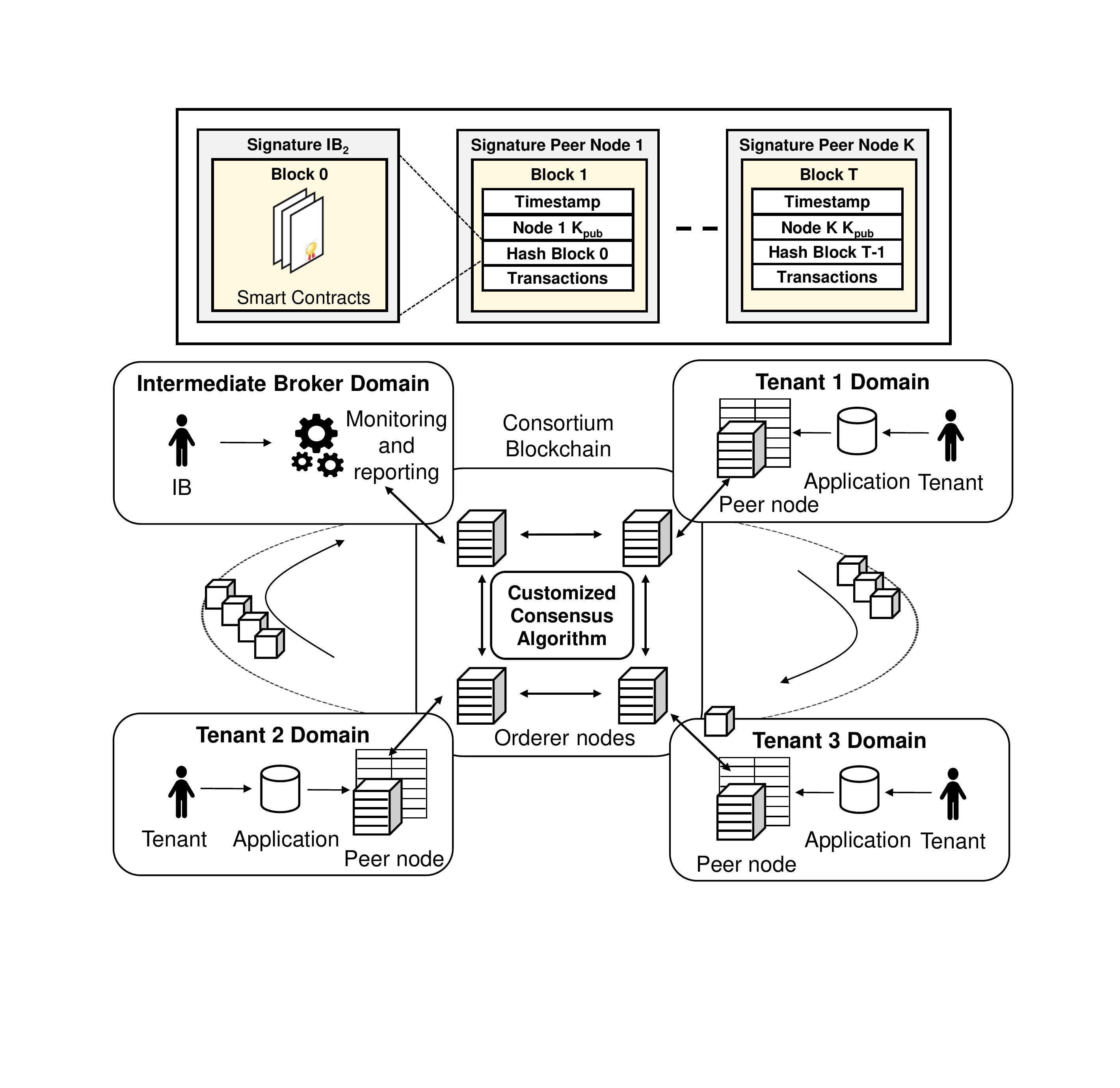}
      \caption{\small Private blockchain architecture supporting network slicing.}
        \label{fig:blockchain_architecture}
\end{figure}

We show the blockchain architecture with involved players supporting network slicing in Fig.~\ref{fig:blockchain_architecture}. Specifically, the IB may join blockchain activities, i.e., it might read the blockchain results, participate to validation and consensus phases as an active member of the blockchain consortium. This implies that the IB can recursively apply confirmed (validated) transactions onto resource scheduling policies that might include (not limited to) RAN/transport and computational resources. 
Despite enabling a more dynamic resource trading market, the blockchain technology would easily allow to keep track of the different resource exchange over time. From the InP perspective, this also simplifies the billing management as each block of transactions stores precise information about the nature of the exchanged resources and the corresponding time window utilization. Moreover, tenants are directly responsible for the management of their requests: once issued, they could not be withdrawn.
Clearly, each IB shows interest in managing properly its resource share with the objective of maximizing the overall final revenue while parsing and processing upcoming slice requests. Let us assume that each tenant $\tau$ can issue multiple slice requests so that the IB can collect all coming slice requests $\Psi_j$ with $j\in\mathcal{J}$. Let us denote $x_{i,j}$ as a decision variable indicating whether $i$-resources of request $j$ is assigned (to the tenant issuing such a request) whereas $y_j$ is used to prevent from partially assigning resources to a single slice request: in other words, a slice request is accommodated only if all types of demanded resources can be assigned to the tenant thereby guaranteeing a correct end-to-end slice instantiation. We can formally write the following optimization problem:

\vspace{2mm}\noindent \textbf{Problem}~\texttt{IB-REVENUE-MAX}:
\begin{equation*}
\label{pr:problem}
\begin{array}{ll}
\text{maximize}   & \sum\limits_{j} \mu_{j} y_{j}       \\
\text{subject to} &  \sum\limits_{j} \pi_i^{(j)} x_{i,j} \leq r_i, \quad\forall i\in\mathcal{I};\\
				  & \sum\limits_{i} x_{i,j} \geq Iy_j, \quad\forall j\in\mathcal{J};\\
				  & \sum\limits_{i} x_{i,j} \leq Iy_j, \quad\forall j\in\mathcal{J};\\
				  &  x_{i,j} \in \{0,1\}, \quad\forall i\in\mathcal{I}, \forall j\in\mathcal{J};\\
				  &  y_{j} \in \{0,1\}, \quad\forall j\in\mathcal{J}.				  
\end{array}
\end{equation*}
where $\mu_j = \sum_i\theta_i^{(j)}, \forall j\in\mathcal{J}$ represents the overall revenue from all types of resources included within the slice request $j$, and $\mathcal{I}$ represents the resource set with $I=|\mathcal{I}|$.
The above optimization problem can be easily mapped onto a integer-linear programming (ILP) problem and solved by means of commercial solvers, e.g.~\cite{cplex}.

\indent \textbf{Consensus algorithm}. The ownership of a resource set can be transferred from one tenant to another by invoking a transaction on the blockchain. The transaction is validated only if all the relevant parties agree, namely, a consensus among peer nodes is reached. When dealing with consensus algorithms, a trade-off between transaction throughput and latency must be considered. We define transaction throughput as the number of transactions that the system can handle per unit of time. In realistic scenarios, this number can range over a wide range depending on the study use-case. For example, public BitCoin's network supports $7$ transactions per second, while the financial networks of MasterCard and Visa handle up to ~$60000$ ~\cite{Luu2016}. Obviously, different consensus algorithms provide different latency. For this reason, we let each IB $b_k$ choose the preferred method according to its service requirements. In general, being \name{} a permissioned framework, we suggest the use of relatively light mechanisms, like Practical Byzantine Fault Tolerance (PBFT) consensus protocol~\cite{PBFT}, Kafka~\cite{Kafka} or Raft~\cite{Raft} to allow fast convergence to a common agreement and speed up the resource exchange process.

\section{Proof-of-Concept Evaluation}
\label{sec:perf_eval}
We implement \name{} on top of Hyperledger Fabric~\cite{HyperledgerFabric}, an open-source framework for developing permissioned blockchains within private enterprises, and make use of its benchmarking tool, namely Hyperledger Caliper, to evaluate the blockchain performance in network slicing scenarios. 

\indent \textbf{Experimental setup.} Our Proof-of-Concept (PoC) architecture consists of $3$ IBs and a variable number of orderer nodes that depends on the adopted consensus algorithm. Such entities run as Docker containers on an Intel Xeon CPU E5-2630v3 32-Core @2.4GHz 64GB RAM shared platform. 

\begin{figure*}[t!]
    \centering
    	  \begin{subfigure}[b]{.48\textwidth}
          \includegraphics[trim = 0cm 0cm 0cm 0cm,clip, width=\linewidth]{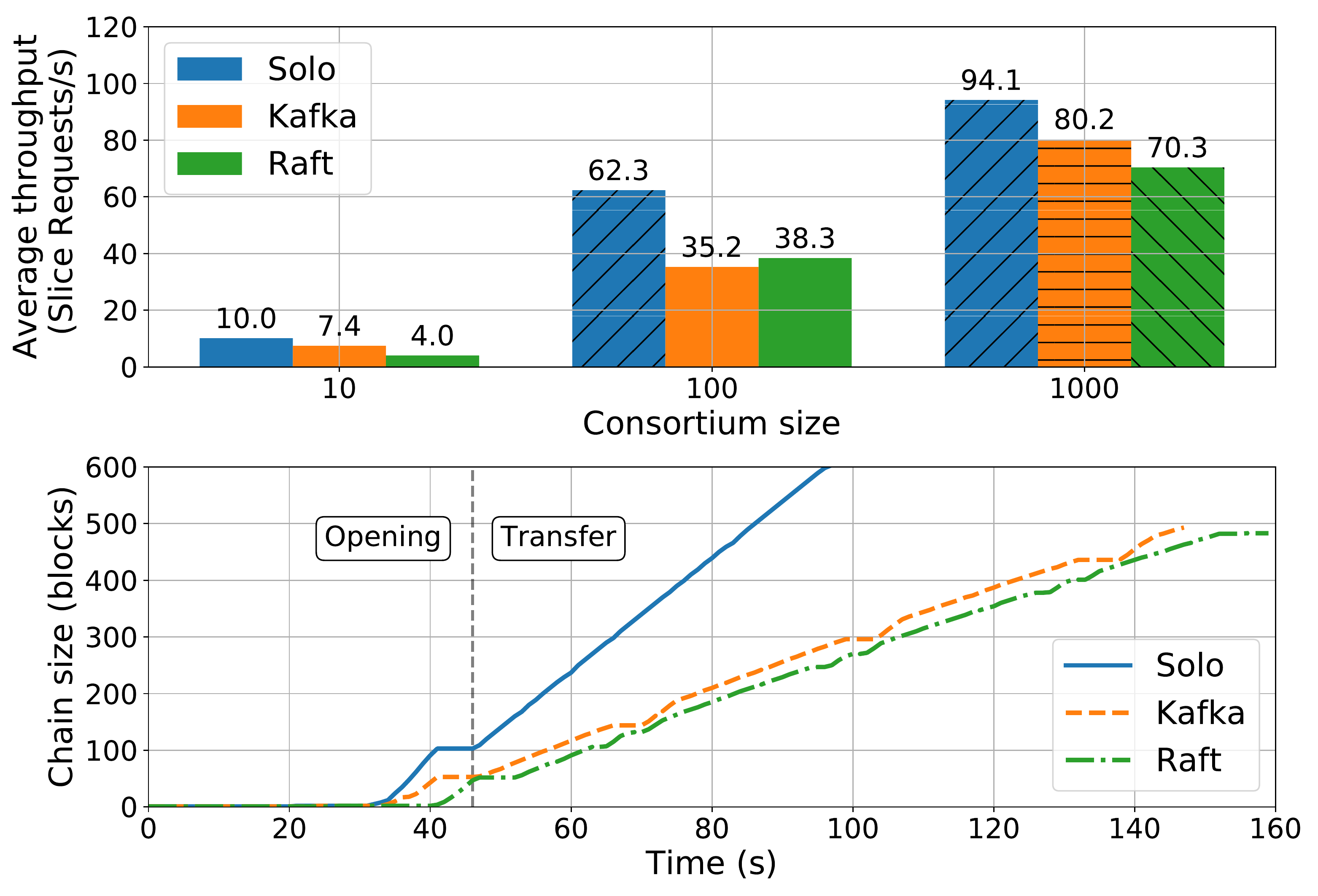}
          \caption{Slice request throughput and blockchain size growth for \\ different consensus algorithms and consortium size. }
          \label{fig:transaction_throughput}
      \end{subfigure}%
	  \begin{subfigure}[b]{.49\textwidth}
          \includegraphics[trim = 0cm 0cm 0cm 0cm,clip, width=\linewidth]{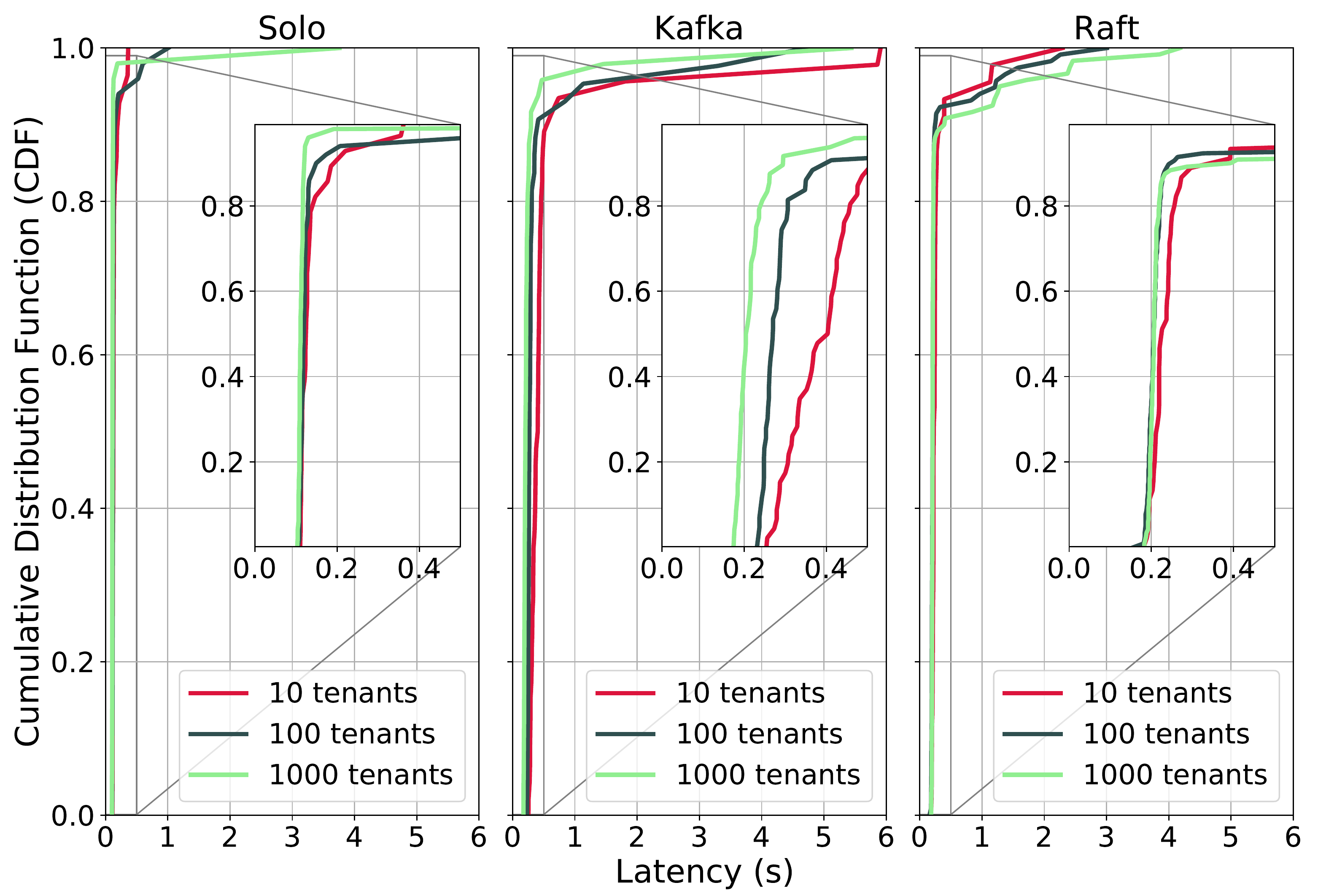}
        \caption{CDF of the slice request validation latency experienced by tenants for different consensus algorithms and consortium size.}
        \label{fig:transaction_latency}
      \end{subfigure}%
    \caption{\small Performance evaluation for different consensus algorithms and consortium size a) Slice request throughput and Blockchain size growth b) Validation latency.}
	\label{fig:sims_results}
\vspace{-4mm}
\end{figure*}

We guarantee the isolation among consortia through the definition of dedicated and encrypted communication channels. Moreover, we set the maximum number of entries per block to $20$ and the block timeout\footnote{We select such values as they maximize the throughput at a minimum latency cost as proved hereafter in the section.} to $300$ ms.
This last metric specifies the amount of time (after receiving the first transaction) each orderer waits before publishing a new set of proposed transactions to other peer nodes.
Please note that the choice of those parameters may strongly affect the blockchain performance. In particular, although decreasing the block timeout improves the latency, setting it to low values may decrease the overall throughput as new blocks would not be filled up to their maximum capacity. To limit the impact of this trade-off on our results, we do not modify these settings throughout this section. 

The benchmark process consists of two phases, dubbed as opening and transfer. In the initial phase, we create tenant instances and assign them with an equal amount of resources such that all available resources at IB side are assigned. Once assigned, each tenant might decide to free or seek additional resources based on a random value drawn from a uniform distribution between $0$ and $30\%$ of the initially assigned amount\footnote{We empirically prove that the choice of this value leads to convergence within a reasonable time.}. During the transfer phase, tenants issue Slice Requests (SRs), modeled as tuples $\Psi_\tau = \{\rho, \eta, \gamma \}$, where $\rho, \eta, \gamma \in \mathcal{R}_k$ represent the percentage of required radio access, transport and core cloud resources, respectively. In case the SR does not fit the availability or the need of the involved tenants (SR collision), it is automatically rejected and the respective transaction is dropped.

\indent \textbf{Full-scale evaluation.} With the first experiment we evaluate the performance of our framework in terms of slice request throughput and latency. We compare two popular consensus algorithms (Kafka and Raft) against a single orderer configuration (Solo) that does not require any consensus process. The top of Fig.~\ref{fig:transaction_throughput} shows the average SR throughput of the platform in the transfer phase for an increasing consortium size and fixed SR rate of $150$ SRs/s to emulate high load conditions. In these settings, especially for a small consortium size, the limiting factor of the blockchain performance throughput is the Multiversion Concurrency Control (MVCC) process. As we issue SRs at a very high rate, the same database entry, e.g. the resources assigned to a specific tenant, may be edited by a new request before the completion of the validation process involving it.
This raises a database inconsistency, dubbed as Read/Write (RW) conflict, which prevents the current transaction to be successful. As shown in the figure, this problem is mitigated by an increasing consortium size.

Fig.~\ref{fig:transaction_latency} depicts the Cumulative Distribution Function (CDF) of the experienced latency by the successful SRs.
As expected, the best latency performance is obtained when no distributed consensus mechanism is in place, i.e., Solo. However, despite being the fastest scheme, this single-node approach is not fault tolerant.
It can be noticed that the transaction exchange and validation process introduce a small time overhead for the Kafka and Raft cases, which however has negligible impact, especially when compared to the onboarding time required e.g., by virtualized infrastructures to setup virtual services~\cite{BSecNFVO}.
The blockchain growth rate is also affected by the different consensus scheme, as shown at the bottom of Fig.~\ref{fig:transaction_throughput}, which refers to the consortium size case of $1000$ tenants. We plot the evolution of the chain size over time and mark the beginning of the transfer phase with a dashed vertical line. It can be noticed that the blockchain grows at a rate proportional to the average throughput since blocks are filled up to their maximum capacity.

\begin{figure}[t!]
      \centering
      \includegraphics[clip, trim= 0cm 0cm 0cm 0cm,   width=0.48\textwidth]{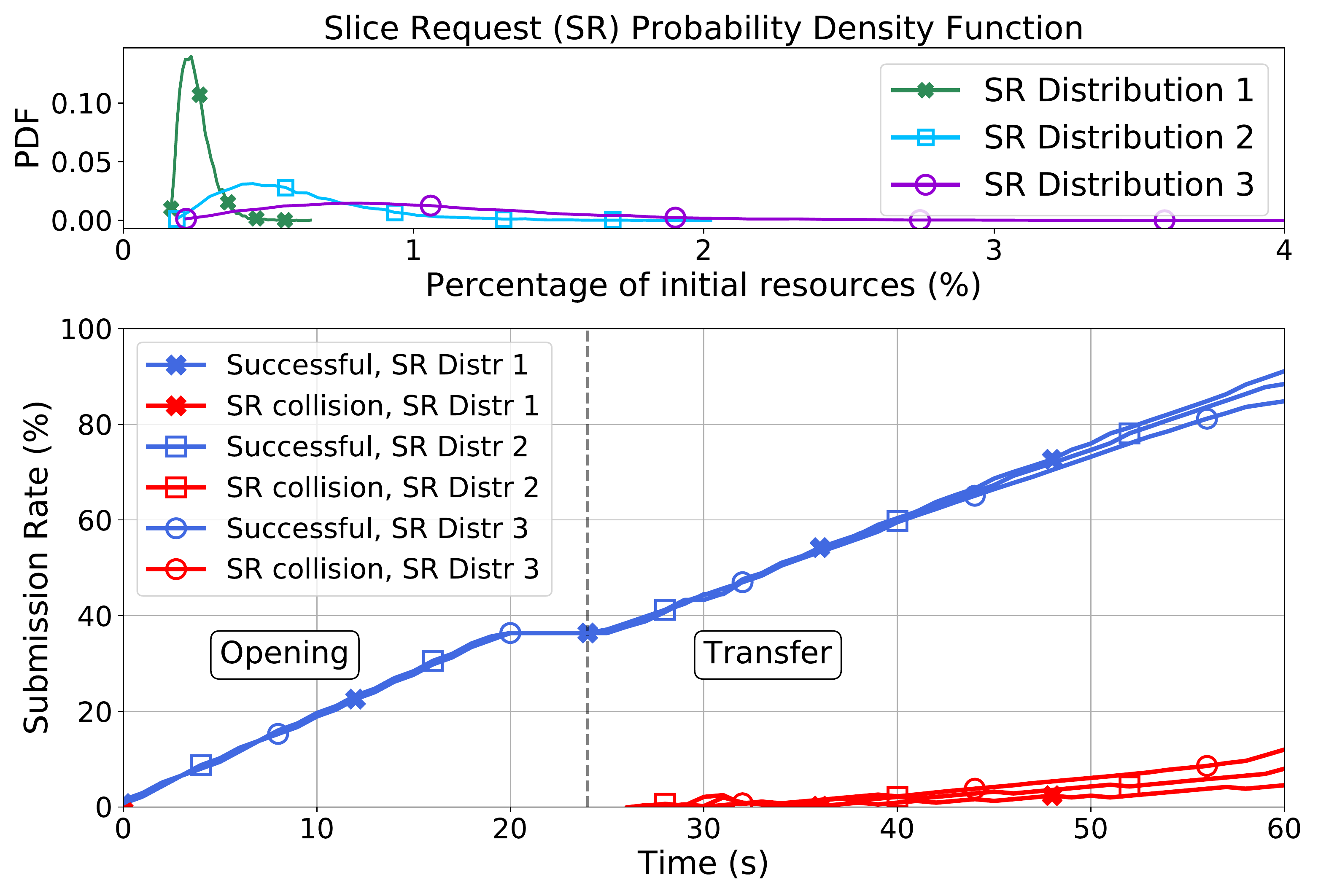}
      \caption{\small Transaction acceptance and error rates for different scenarios.}
        \label{fig:acceptance_rate}
\end{figure}

\indent \textbf{Brokering scenario evaluation.}
The second experiment focuses on evaluating the capabilities of the system when dealing with the brokering scenario. To this aim, we consider $3$ IBs managing a consortium of $1000$ tenants, correspondingly. In light of the performances shown above, we select Kafka as consensus algorithm for its high fault-tolerance and scalability~\cite{Performance_Benchmarking}. We assume that resource request values $\rho, \eta, \gamma$ are drawn from a right-skewed distribution over a positive interval as resource requests must be non-negative. Such distributions are depicted at the top of Fig.~\ref{fig:acceptance_rate} for different demand ranges, spanning from $0.1$\% to $4$\% of the tenant initial resources. Note that since we assume the same distribution for all resource requests within the same slice, it is dubbed as SR PDF.

The bottom of Fig.~\ref{fig:acceptance_rate} illustrates the system behavior for a constant submission rate of $50$ SRs/s so as to keep RW Conflicts to a minimum (around $2\%$ of the submitted SRs). In such operational conditions, errors raise only in case of SR collisions. It is worth noting that SR collision rate increases along with the SR variance. Specifically, SR distributions with high variance leads to tenant satisfaction more quickly than with a lower variance. Additionally, the closer to tenant satisfaction, the lower the resource availability and, in turn, the smaller the likelihood of a request to be accepted by the system.
\section{Conclusion}
\label{sec:conclusion}

Network slicing has been identified as a key enabler for the development of novel business models in 5G and beyond mobile networks. In this paper we introduced a hierarchical blockchain-based framework, \name{}, that provides a brokering solution between the infrastructure provider and network tenants willing to pay for acquiring, exchanging and managing network and computational slice resources within the domain of an intermediate broker.
We developed a Proof-of-Concept implementation leveraging on the open-source Hyperledger platform and showed that our approach is feasible and scalable (up to $1000$ tenants were considered) with different state-of-the-art consensus algorithms.

\section*{Acknowledgment}
The research leading to these results has been partially supported by the H$2020$ MonB5G Project under grant agreement number $871780$.

\bibliographystyle{IEEEtran}
\bibliography{References}

\end{document}